\documentclass[a4paper, 11pt]{article}

\usepackage[a4paper, bindingoffset=0.2in,%
            left=1.0in, right=1.0in, top=1.0in, bottom=0.9in,%
            footskip=.25in]{geometry}
            
\usepackage[english]{babel}
\usepackage[utf8x]{inputenc}
\usepackage[T1]{fontenc}

\usepackage[numbers,square]{natbib}

\usepackage{multicol}
\usepackage{multirow}

\usepackage[belowskip=-1pt, aboveskip=3pt]{caption}

\usepackage{titlesec}
\titlespacing\section{0pt}{12pt plus 2pt minus 2pt}{2pt plus 0pt minus 0pt}
\titlespacing\subsection{0pt}{12pt plus 1pt minus 1pt}{0.5pt plus 0.5pt minus 0.5pt}
\titlespacing\subsubsection{0pt}{0pt plus 0pt minus 0pt}{0pt plus 0pt minus 0pt}

\usepackage[colorlinks=true, allcolors=blue]{hyperref}

\usepackage{booktabs}

\usepackage{graphicx}
\usepackage{float}
\usepackage{pdfpages}

\usepackage{xurl}

\usepackage{fancyhdr}
\fancypagestyle{firstpagestyle}{%
  \fancyhf{}
  \rfoot{\thepage}
  
}

\def\BibTeX{{\rm B\kern-.05em{\sc i\kern-.025em b}\kern-.08em
    T\kern-.1667em\lower.7ex\hbox{E}\kern-.125emX}}

\title{Global Built-up and Population Datasets:\\Which ones should you use for India?}
\makeatletter
\renewcommand\@date{{%
  \large\centering
  \begin{tabular}{@{}c@{}}
    Pratyush Tripathy \\
    \normalsize \href{mailto:pratyush@iihs.ac.in}{pratyush@iihs.ac.in}
  \end{tabular}%
  \quad\quad\quad\quad
  \begin{tabular}{@{}c@{}}
    Krishnachandran Balakrishnan \\
    \normalsize \href{mailto:kbalakrishnan@iihs.co.in}{kbalakrishnan@iihs.co.in}
  \end{tabular}

  \bigskip

  \em{\small Geospatial Lab, Indian Institute for Human Settlements, Bengaluru, India. 560 080}

  \bigskip

}}
\makeatother

\hypersetup{
pdftitle={IDC-Gridded-Population-India},
pdfsubject={Land-Cover-And-Gridded-Population},
pdfauthor={Pratyush Tripathy, Krishna Balakrishnan},
pdfkeywords={Land Cover, Gridded Population, SDGs},
}

\begin{document}

\maketitle
\thispagestyle{firstpagestyle}


\begin{multicols}{2}

\section*{Abstract}

Multiple global land cover and population distribution datasets are currently available in the public domain. Given the differences between these datasets and the possibility that their accuracy may vary across countries, it is imperative that users have clear guidance on which datasets are appropriate for specific settings and objectives. Here we assess the accuracy of three global 10m resolution built-up datasets (ESRI, GHS-BUILT-S2 and WSF) and three population distribution datasets (HRSL 30m, WorldPop 100m, GHS-POP 250m) for India. Among built-up datasets, the GHS-BUILT-S2 is the most suitable for India for the 2015-2020 time period. To assess accuracy of population distribution datasets we use data from the 2011 Census of India at the level of 37,137 village and town polygons for the state of Bihar in India. Among the global datasets, HRSL has the best results. We also compute error metrics for IDC-POP, a 30m resolution population dataset generated by us at the Indian Institute for Human Settlements. For Bihar, IDC-POP outperforms all three global datasets.

\section{Introduction}
Since 2015, international framework agreements like the Sustainable Development Goals, Sendai Framework for Disaster Risk Reduction and Habitat III New Urban Agenda \cite{un2015, unisdr2015, un2017} have been shaping global development priorities. But effective implementation and monitoring of progress towards these require data at sufficient temporal and spatial resolution \cite{un2015, ieag2014}. Geospatial datasets like land cover and population distribution maps are crucial for monitoring numerous indicators related to these goals \cite{trends2020} and identifying populations that are exposed to various hazards \cite{feng2020, kavvada2020, kuffer2020, qui2019}.

To address this issue, since 2015, several organizations have released land cover data at 10m resolution and population datasets at 250m or finer resolution. But as multiple such datasets become available in the public domain, it is important that users have clear guidance in terms of the accuracy and suitability of these datasets for specific applications. Moreover, accuracy may vary across countries. Here we compare some of these recently released global datasets to identify the most appropriate ones for mapping built-up and population distribution in India.

\section{Built-up Land Cover}
In June 2021, ESRI announced the release of its 10m resolution, 10 class, Global Land Cover Map, generated in partnership with Impact Observatory \cite{esri2021}. It uses a land cover classification model trained by the Impact Observatory on a dataset developed by the National Geographic Society. The model was applied to Sentinel-2 data for 2020 using the Microsoft Planetary Computer. 

10m resolution land cover data is a significant improvement over other recently released datasets which are mostly at 100m or 30m resolution \cite{buchhorn2020}. The Global Human Settlement Layer (GHS-BUILT-S2) derived from Sentinel-2 by the European Commission Joint Research Center (JRC) and the World Settlement Footprint (WSF) generated by the German Aerospace Center (Deutsches Zentrum für Luft- und Raumfahrt or DLR) are the only other 10m global open datasets available \cite{corbane2021, marconcini2020}. Of these, the GHS-BUILT-S2 is a probabilistic 10m built-up layer and the WSF is a combination of 10m and 30m resolution cells derived from Sentinel-1 and Landsat 8 data. But both of these datasets have only one land cover class (built-up) and are available only for one time period -- for 2018 and 2015 respectively. In comparison, the 10 class ESRI land cover data is meant to be updated annually. 

\end{multicols}

\begin{figure*}[ht]
\centering
\includegraphics[width=15 cm]{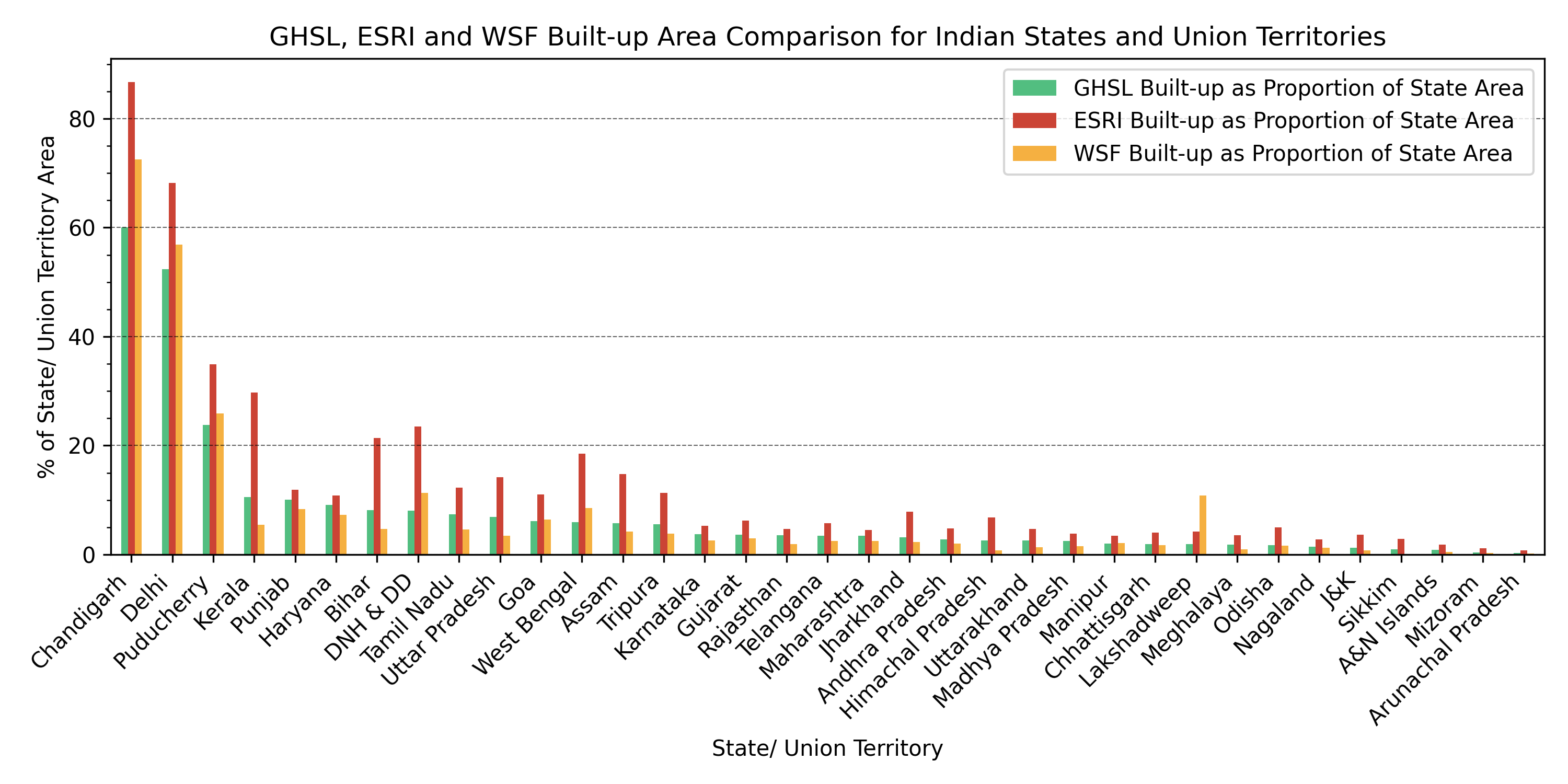}
\caption{Comparison of built-up areas (as \% of state/UT area) of Indian States and Union Territories computed using GHSL and ESRI 10m built up layers.}
\label{fig:fig1}
\end{figure*}

\vspace{-0.25cm}
\begin{multicols}{2}

\subsection{State level built-up comparison}
In Figure \ref{fig:fig1} we compare percentage built-up area estimates for all states and union territories of India derived from the ESRI (2020), GHS-BUILT-S2 (2018) and WSF (2015) datasets. We focus on these three since these are the highest resolution global land cover datasets currently available. We evaluate over and under prediction with respect to GHS-BUILT-S2 since it falls in between the ESRI and WSF layers in terms of temporal sequence with ESRI data being 2 years newer and WSF being 3 years older than GHS-BUILT-S2. Since GHS-BUILT-S2 is a probabilistic layer, we convert it to binary layer for comparison purposes. For this, we extract all cells with probability value greater than or equal to 20\% as suggested by Corbane et al. \cite{corbane2021}.

As evident from Figure \ref{fig:fig1}, in comparison to the GHS-BUILT-S2, the ESRI dataset consistently overpredicts and WSF mostly underpredicts built-up area of states and union territories. In some states like Kerala, Bihar, West Bengal, Assam and Jharkhand, ESRI built-up area is more than twice of GHS-BUILT-S2 and WSF.

In the case of WSF, there is over prediction only in some of the union territories like Lakshadweep, Puducherry, Dadra and Nagar Haveli \& Daman and Diu, and densely built areas like Delhi and Chandigarh. Among larger states WSF overpredicts built-up area only for West Bengal.

\begin{figure*}[ht]
\centering
\includegraphics[width=15cm]{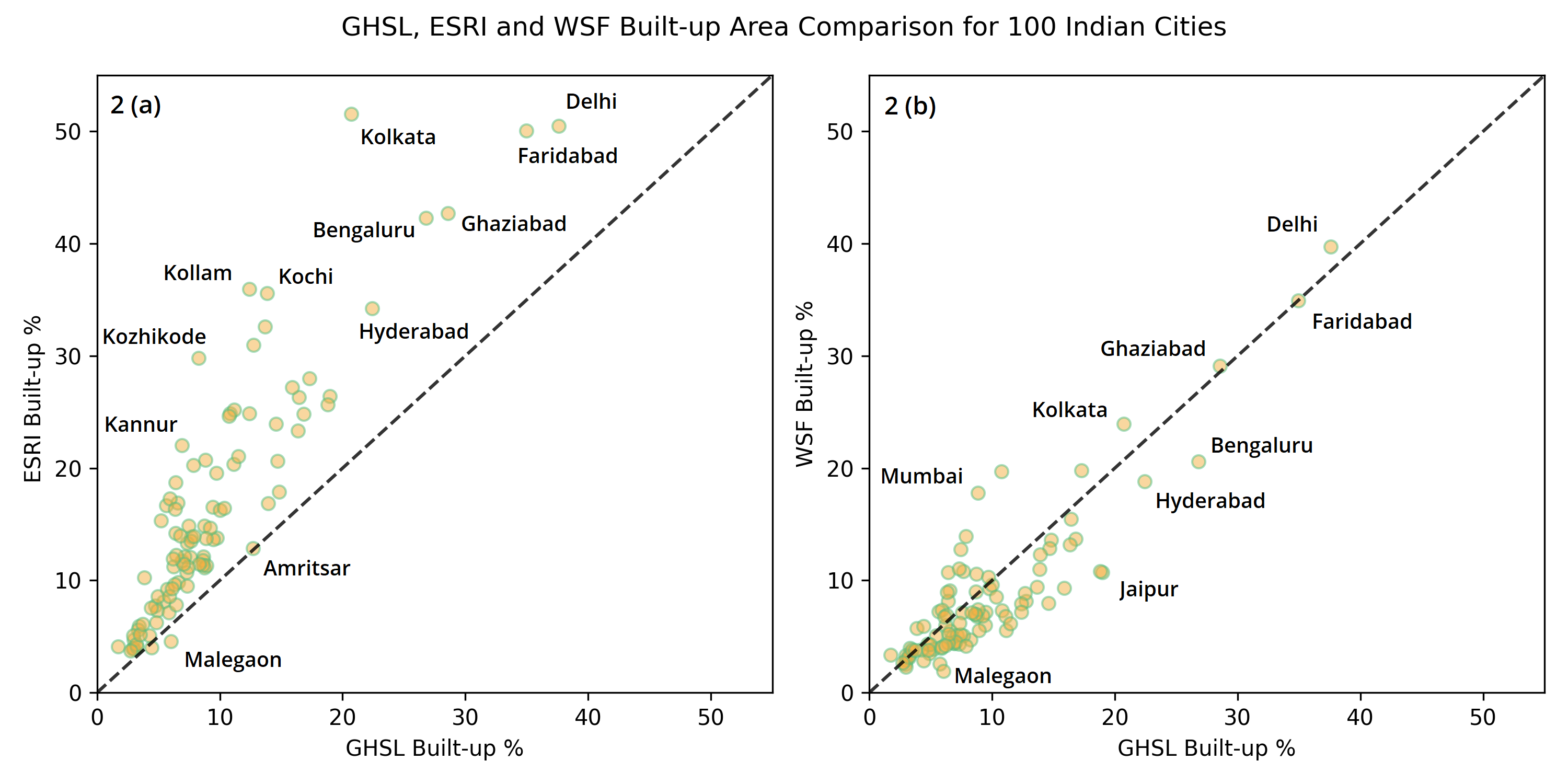}
\caption{Built-up percentage scatter plots for top 100 Indian cities.}
\label{fig:fig2}
\end{figure*}

\subsection{Urban built-up comparison}

For urban areas, we did a similar analysis for the top 100 most populous Indian cities using a 60 km by 60 km area of interest (AOI) centered at the city center \cite{malladi2017}. Figure \ref{fig:fig2} shows the scatter plots of the built-up area percentage derived using ESRI and GHS-BUILT-S2 (Fig. \ref{fig:fig2}a) and WSF and GHS-BUILT-S2 (Fig. \ref{fig:fig2}b).

\newpage

From Figure \ref{fig:fig2}a, it is clear that except for some of the smaller cities, compared to GHS-BUILT-S2, the ESRI dataset significantly overpredicts built-up area in cities also. In comparison, Figure \ref{fig:fig2}b shows that while there is better agreement between WSF and GHS-BUILT-S2 built-up maps for cities, like in the case of states, WSF is once again slightly under predicting in relation to GHS-BUILT-S2.

\subsection{Visual comparison of built-up layers and Sentinel-2 imagery}

\begin{figure*}[p]
\noindent\centering
\resizebox{0.8\textwidth}{!}{\includegraphics{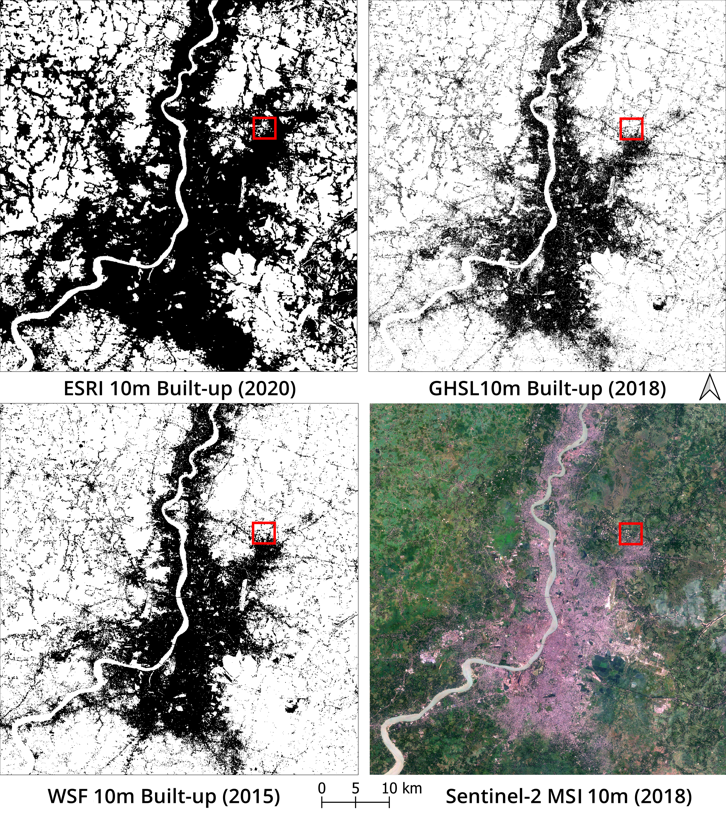}}
\caption{ESRI built-up layer, GHS-BUILT-S2, WSF and Sentinel-2 MSI for Kolkata}
\label{fig:fig3}

\resizebox{0.8\textwidth}{!}{\includegraphics{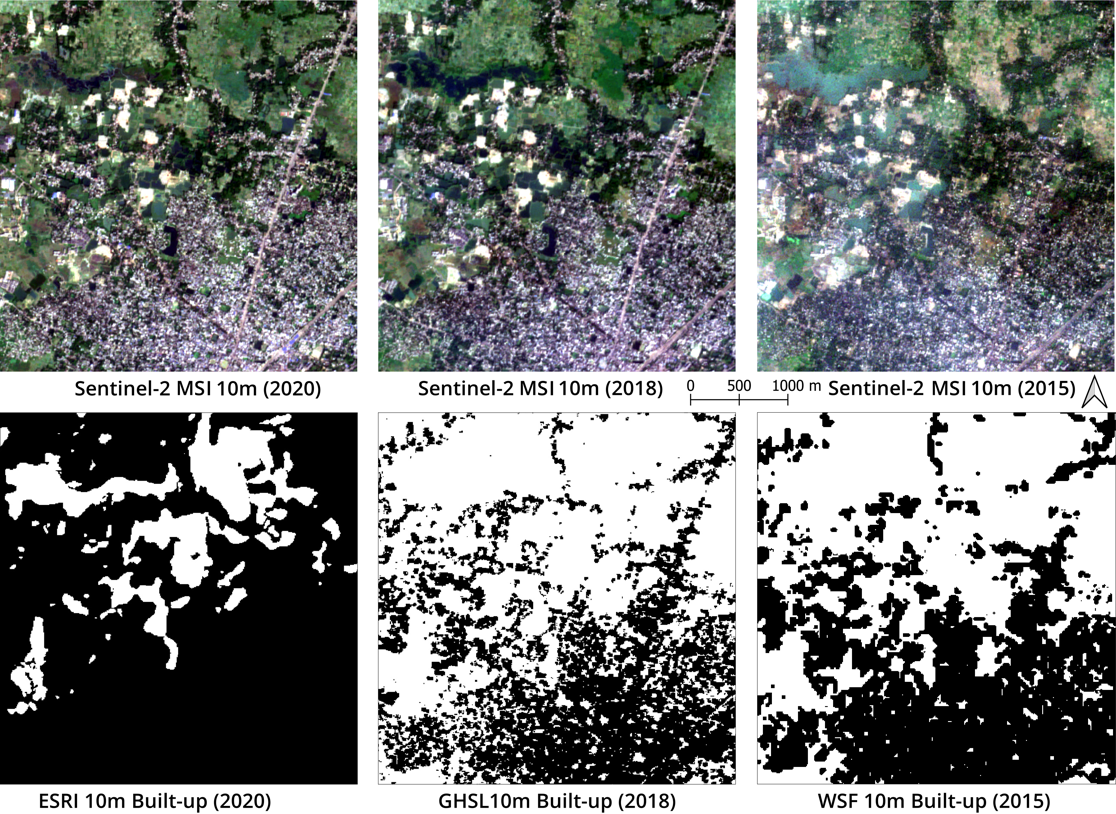}}
\caption{ESRI built-up layer, GHS-BUILT-S2 and WSF along with Sentinel-2 MSI of corresponding years for area within the red rectangle in Fig. \ref{fig:fig3}}
\label{fig:fig4}
\end{figure*}

Since there is no reference built-up layer for India which can be used to assess the accuracy of these global datasets, the only way to identify which of these datasets may be closer to reality is through visual inspection of the built-up maps and satellite imagery. Figures \ref{fig:fig3} and \ref{fig:fig4} compare all the three built-up datasets and Sentinel-2 satellite imagery for the city of Kolkata. To account for the temporal differences between datasets, in Figure \ref{fig:fig4} we compare each built-up dataset with Sentinel-2 MSI 10m imagery for the corresponding year.

\begin{figure*}[p]
\centering
\resizebox{0.8\textwidth}{!}{\includegraphics{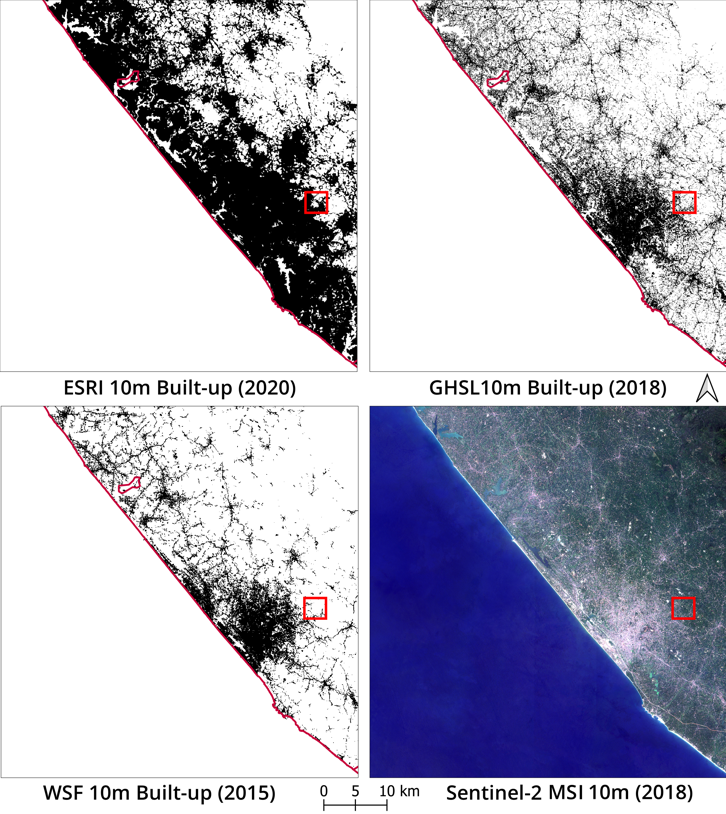}}
\caption{ESRI built-up layer, GHS-BUILT-S2, WSF and Sentinel-2 MSI for Trivandrum}
\label{fig:fig5}

\resizebox{0.8\textwidth}{!}{\includegraphics{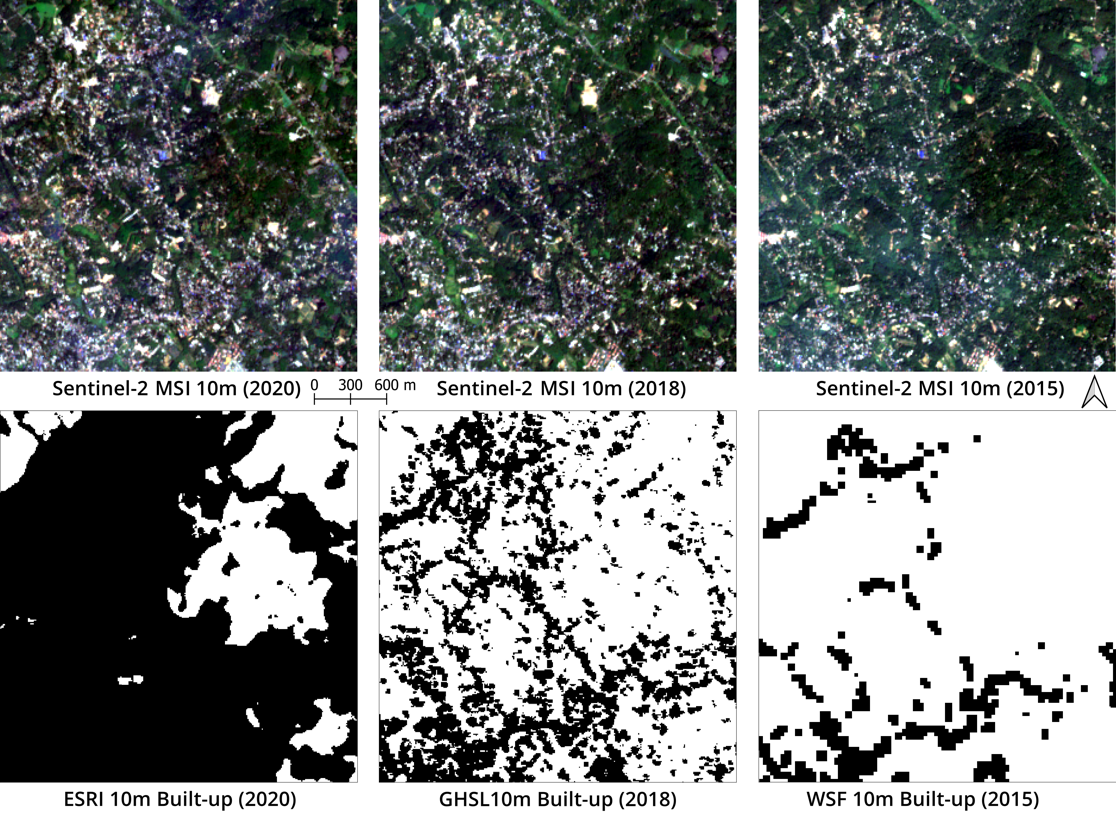}}
\caption{ESRI built-up layer, GHS-BUILT-S2 and WSF along with Sentinel-2 MSI of corresponding years for area within the red rectangle in Fig. \ref{fig:fig5}}
\label{fig:fig6}
\end{figure*}

From a visual inspection of Figures \ref{fig:fig3} and \ref{fig:fig4}, we can see that GHS-BUILT-S2 matches closely with the built-up extent that Sentinel-2 from the corresponding year appears to show. On the other hand, the ESRI built-up layer shows considerable overprediction compared to the built-up extent visible in the Sentinel-2 imagery for 2020. WSF misses some of the finer built-up patches towards the periphery, but for the main built-up extent it seems to be overpredicting compared to GHS-BUILT-S2 and 2015 Sentinel-2 imagery. This could be a result of WSF relying partly on 30m raw landsat data for these areas. Overall, GHS-BUILT-S2 seems to be the most accurate built-up layer for the 2015 to 2020 time period. In Figure \ref{fig:fig5}, we perform the same comparison for Trivandrum since it is a much smaller city from a very different part of the country.

From Figures \ref{fig:fig5} \& \ref{fig:fig6} it is clear that the ESRI dataset substantially overpredicts built-up. Once again WSF misses the finer built-up patches, perhaps due to the partial reliance on 30m raw landsat data. As evident from Figure \ref{fig:fig6}, the temporal differences do not explain the over and under prediction of ESRI and WSF datasets. Once again for the period from 2015 to 2020 GHS-BUILT-S2 is the most reliable built-up layer, although with respect to 2015 Sentinel imagery it is overpredicting built-up since GHS-BUILT-S2 was generated using 2018 imagery. However, considering the coarseness of WSF and the extent of underprediction for 2015 in WSF, we conclude that it is better to use GHS-BUILT-S2 for 2015 also.

\end{multicols}

\begin{multicols}{2}

Below we compare state level built-up maps generated from all the three datasets for Kerala and Bihar. As seen in Figure \ref{fig:fig1}, for both these states there is considerable variation across datasets with the ESRI prediction being roughly 150\% more than GHS-BUILT-S2 and WSF prediction being roughly 50\% less than GHS-BUILT-S2.

The ESRI layer shows Kerala as having dense and nearly contiguous built-up across the entire state--which is clearly a very significant overprediction if one compares it with freely available Google Earth imagery. Compared to GHS-BUILT-S2, WSF visibly predicts less built-up area at the state level also. But based on Figures \ref{fig:fig4} and \ref{fig:fig6}, it seems likely that GHS-BUILT-S2 is closer to reality for the time period from 2015 to 2020.

\end{multicols}

\vspace{-0.5cm}
\begin{figure}[H]
\centering
\resizebox{0.88\textwidth}{!}{\includegraphics{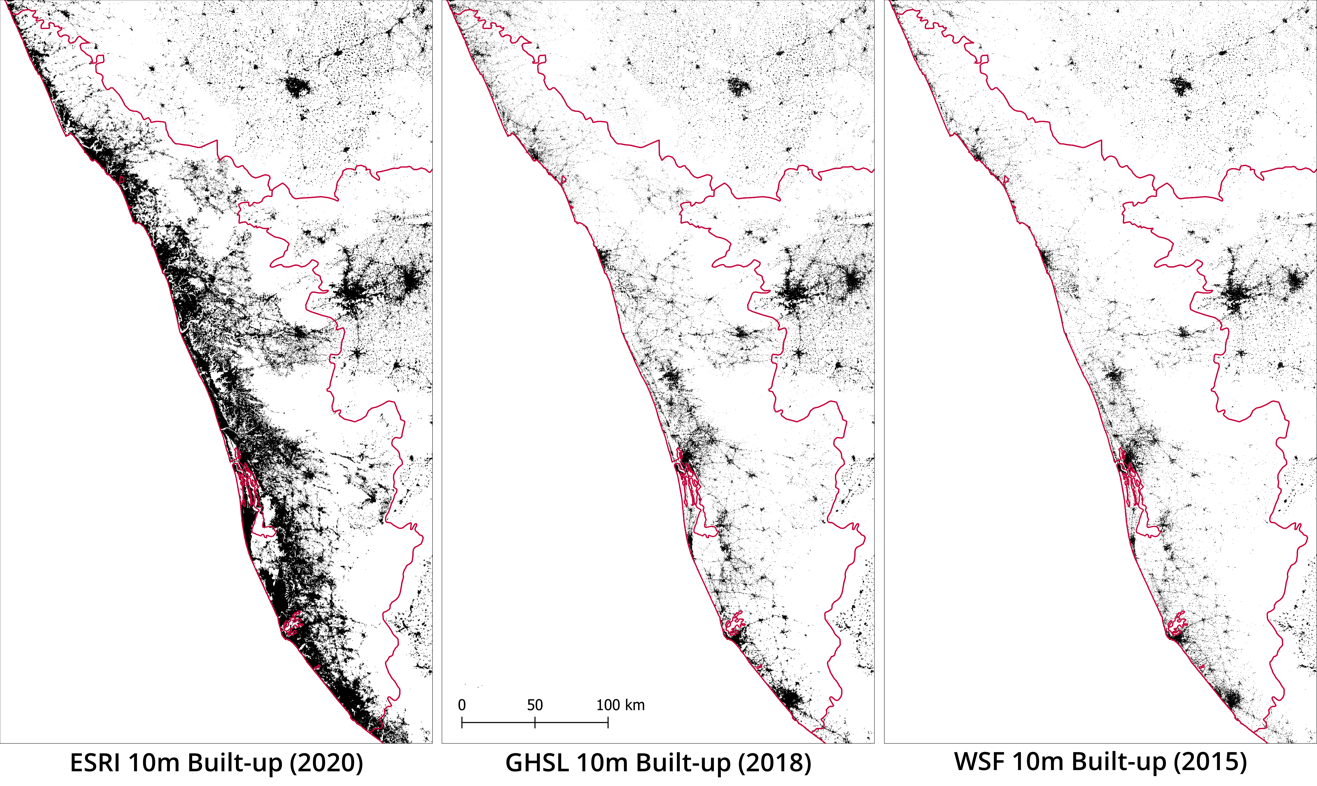}}
\caption{Binary built-up layers from ESRI, GHS-BUILT-S2 and WSF for Kerala.}
\label{fig:fig7}
 
\vspace{0.5cm}
\resizebox{0.88\textwidth}{!}{\includegraphics[width=13.5 cm]{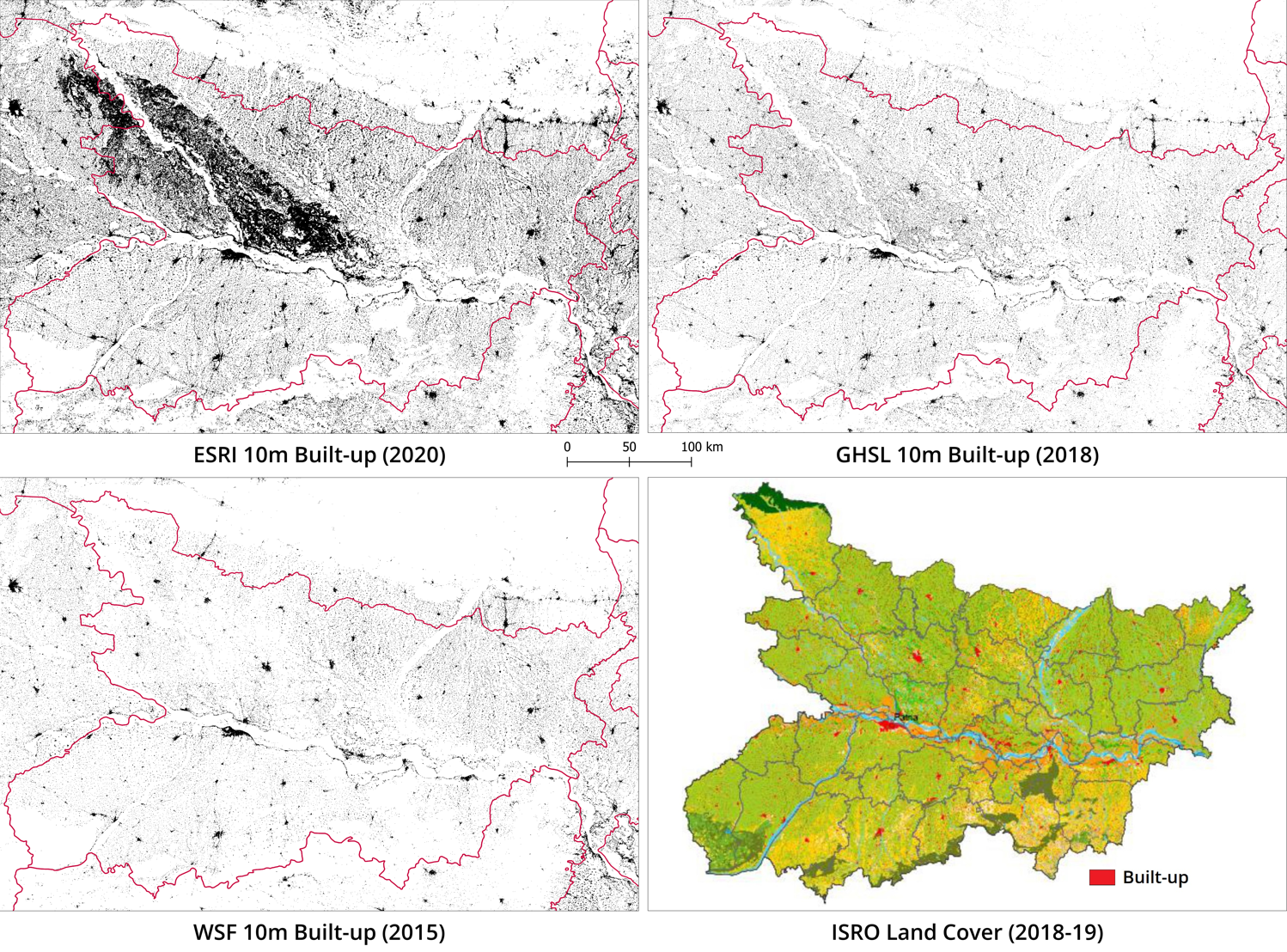}}
\caption{Comparison of binary built-up layers from ESRI, GHS-BUILT-S2 and WSF with ISRO 2018-2019 land cover data for Bihar, India. \cite{nrsc2020}}
\label{fig:fig8}
\end{figure}

\begin{multicols}{2}

In the ESRI dataset we see considerable overprediction in the built-up layer for Bihar too. In Figure \ref{fig:fig8} we provide land cover data of Bihar generated by the Indian Space Research Organisation (ISRO) for the year 2018-19 as reference \cite{nrsc2020}. The built-up area in ISRO’s land cover layer is clearly much closer to GHS-BUILT-S2 and WSF datasets. In particular, in the ESRI layer there is a very striking patch of dense built-up seen to the north-west part of the state which is missing from all other datasets. Examination of Google Earth imagery also shows that this is clearly erroneous.

\section{Gridded Population}
As of 2021, multiple global gridded population datasets are available. SDSN TReNDS provides a detailed comparison of seven such datasets and guidelines for usage in their recent report \cite{trends2020}. Of these seven, five are freely available to the public, and of these, we focus on the three highest resolution datasets: High Resolution Settlement Layer (HRSL, 2015, 30m), WorldPop (2011, 100m) and Global Human Settlement - Population (GHS-POP, 2015, 250m) \cite{fb2016, ghsl2016, worldpop2018}. The analysis below is restricted to these three datasets for India and is intended to supplement the analysis and guidance provided in the SDSN TReNDS report.

The HRSL dataset has been developed through a collaboration between Facebook, Center for International Earth Science Information Network at Columbia University (CIESIN) and The World Bank, while CIESIN and the JRC developed the GHS-POP dataset. Both of these datasets use 2015 population numbers from the CIESIN Gridded Population of the World version 4.10 (GPWv4.10) \cite{gpw410a}. The WorldPop dataset has been generated by the WorldPop team based at the University of Southampton, and uses 2011 census data. The input population for all three datasets are at the level of 5,967 sub-districts of India.  

\subsection{Accuracy assessment method}
Accuracy assessment of these datasets is typically not possible at the town/village level for India. This is because, for the 2011 Census, there is no publicly accessible, national level village/town boundary dataset with village/town codes to which census population data can be appended.

To address this issue we downloaded publicly available 2011 village boundaries for the state of Bihar from multiple online sources including various github repositories. We then identified the best boundaries based on their correspondence with features like roads, highways and rivers. These boundaries were then cleaned and assigned village/town identifiers using location information from OpenStreetMap and 2001 village/town boundaries--available from Socio-economic Data and Applications Center (SEDAC) hosted at CIESIN. To do this we also matched names of 2001 and 2011 settlements using the Census of India town and village tables. Our final cleaned dataset has 37,137 polygons, of which 36,960 are villages and 177 are towns (census towns + statutory towns). As per the Census of India, Bihar has 44,874 villages and 199 towns with a total population of 104.1 million--roughly the same as Egypt. Although our dataset is still missing 7,936 polygons, we are able to account for 99.5 million people, which is more than 95\% of the total population.

To complete the accuracy assessment, we also needed to ensure that all datasets are mapping 2011 population. As discussed earlier, since HRSL and GHS-POP use GPWv4.10, they are mapping 2015 population. GPWv4.10 estimates 2015 population by applying a district level multiplier to 2011 Census data \cite{gpw410b}. To adjust for this and ensure comparability across datasets, using district boundaries for Bihar, we first calculated 2015 district population from HRSL and GHS-POP. Dividing this 2015 district population estimate by the district population totals from 2011 Census we computed the district level multiplier used in GPWv4.10. HRSL and GHS-POP yielded slightly different multipliers for each district. This may be because of the differences in the input district or sub-district boundaries used by HRSL and GHS-POP. So for each dataset we used the respective multipliers as an adjustment factor to all cells within a district to generate adjusted 2011 gridded population datasets for HRSL and GHS-POP. 

We then used the 37,137 village/town polygons layer to assess the accuracy of the adjusted HRSL and GHS-POP datasets and WorldPop. In addition, using the same approach, we also assessed the accuracy of our own population mapping efforts which we have been carrying out as part of the India Data Cube (IDC) and PEAK-Urban projects at the Indian Institute for Human Settlements (IIHS). Our gridded population dataset (IDC-POP) is of 30m resolution and is generated using a Geographically Weighted Regression based approach as described in \cite{iihs2021}. The GHS-BUILT-S2 10m built-up layer and sub-district level population are the primary inputs for this. It is important to note that for Bihar, at present, we have used the 2018 GHS-BUILT-S2 data to map 2011 population. In our future work we will be using 2011 Landsat data to achieve the same objective. 

\subsection{Accuracy assessment results}

\begin{figure}[H]
\centering
\resizebox{0.95\textwidth}{!}{\includegraphics{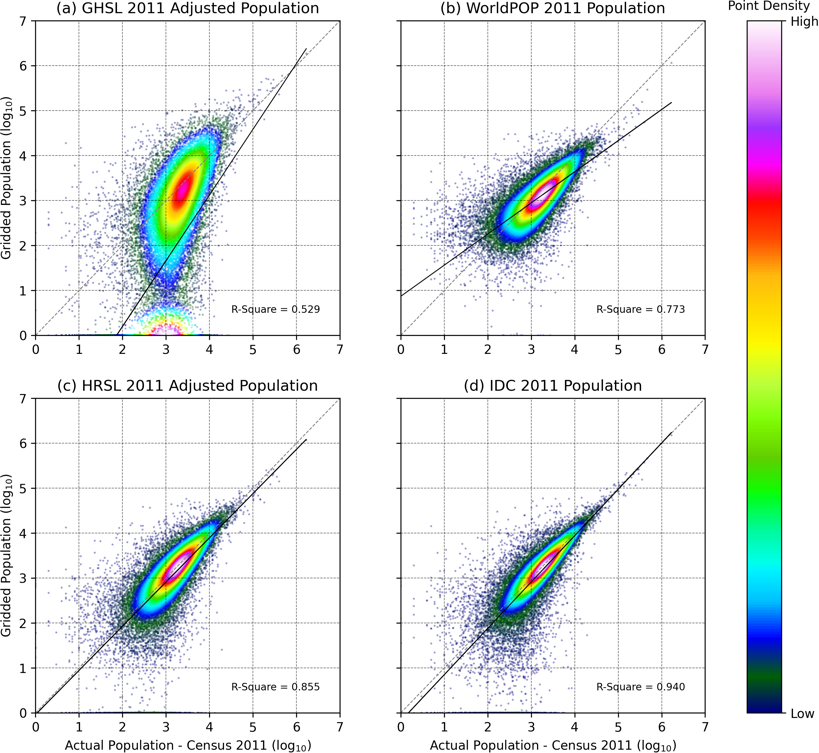}}
\caption{Scatter plot between actual village/town population and (a) GHS-POP, (b) WorldPop, (c) HRSL and (d) IDC-POP.}
\label{fig:fig9}
\end{figure}

Figure \ref{fig:fig9} shows the scatter plot between the actual population and population aggregated from various gridded datasets for 37,137 villages/towns of Bihar. From the scatterplots and R-square values we can see that HRSL is the best among global datasets. Although the overall spread of points for IDC-POP and HRSL scatter plots are similar, from the point density color coding and from R-square values we can see that IDC-POP is performing considerably better than HRSL.

Table \ref{tab:tab1} lists Root Mean Square Error (RMSE), Mean Absolute Error (MAE), Mean Absolute Percentage Error (MAPE), Mean Percentage Error (MPE) and R-square values for all four datasets--first for all 37,137 polygons and then separately for towns (Census towns + Statutory towns) and villages. 

\end{multicols}

\vspace{-0.25cm}

\begin{multicols}{2}

\begin{table}[H]
\centering
\caption{Accuracy metrics for population datasets}
\setlength{\tabcolsep}{0.4\tabcolsep}
\label{tab:tab1}
\resizebox{7.5cm}{!}{%
\begin{tabular}{ *{8}{c} }
    \toprule
    
    \textbf{Dataset} & \textbf{Resol.} & \textbf{Level} & \textbf{RMSE} & \textbf{MAE} & \textbf{MAPE} & \textbf{MPE} & \textbf{R\textsuperscript{2}} \\
    \midrule
    
    \multirow{3}{*}{GHS-POP} & \multirow{3}{*}{250m} & Overall & 7625 & 2477 & 177 & 51 & 0.53 \\
    & & Town & 76845 & 48670 & 144 & 125 & 0.71 \\
    & & Village & 5490 & 2256 & 177 & 51 & -2.24 \\
    \midrule
    
    \multirow{3}{*}{WorldPop} & \multirow{3}{*}{100m} & Overall & 5298 & 1233 & 159 & 104 & 0.77 \\
    & & Town & 71232 & 35817 & 59 & -54 & 0.75 \\
    & & Village & 1975 & 1067 & 160 & 105 & 0.58 \\
    \midrule
    
    \multirow{3}{*}{HRSL} & \multirow{3}{*}{30m} & Overall & 4223 & 1144 & 105 & 61 & 0.86 \\
    & & Town & 52215 & 19683 & 41 & -0.97 & 0.87 \\
    & & Village & 2205 & 1055 & 105 & 61 & 0.48 \\
    \midrule
    
    \multirow{3}{*}{IDC-POP} & \multirow{3}{*}{30m} & Overall & 2710 & 864 & 96 & 58 & 0.94 \\
    & & Town & 29649 & 14378 & 40 & 18 & 0.96 \\
    & & Village & 1780 & 799 & 96 & 58 & 0.66 \\
    \bottomrule
    
\end{tabular}}
\end{table}

For HRSL and GHS-POP, we computed all these metrics for the unadjusted 2015 gridded datasets also. We found that the unadjusted HRSL and GHS-POP had considerably higher error across all metrics. Hence we report only on the results obtained using the adjusted versions of HRSL and GHS-POP which represent 2011 population.

Between the three global gridded datasets, HRSL is the most accurate based on almost all metrics. WorldPop is the next best, and is better than HRSL for villages, based on some metrics (RMSE and R-square). GHS-POP has the highest levels of error and lowest R-square values. GHS-POP is particularly poor at predicting village populations as seen from the negative R-square value. In comparison, IDC-POP outperformed all the three global datasets on almost all error metrics. HRSL performs better than IDC-POP only in the case of MPE for towns.

\begin{figure}[H]
\centering
\resizebox{0.95\textwidth}{!}{\includegraphics{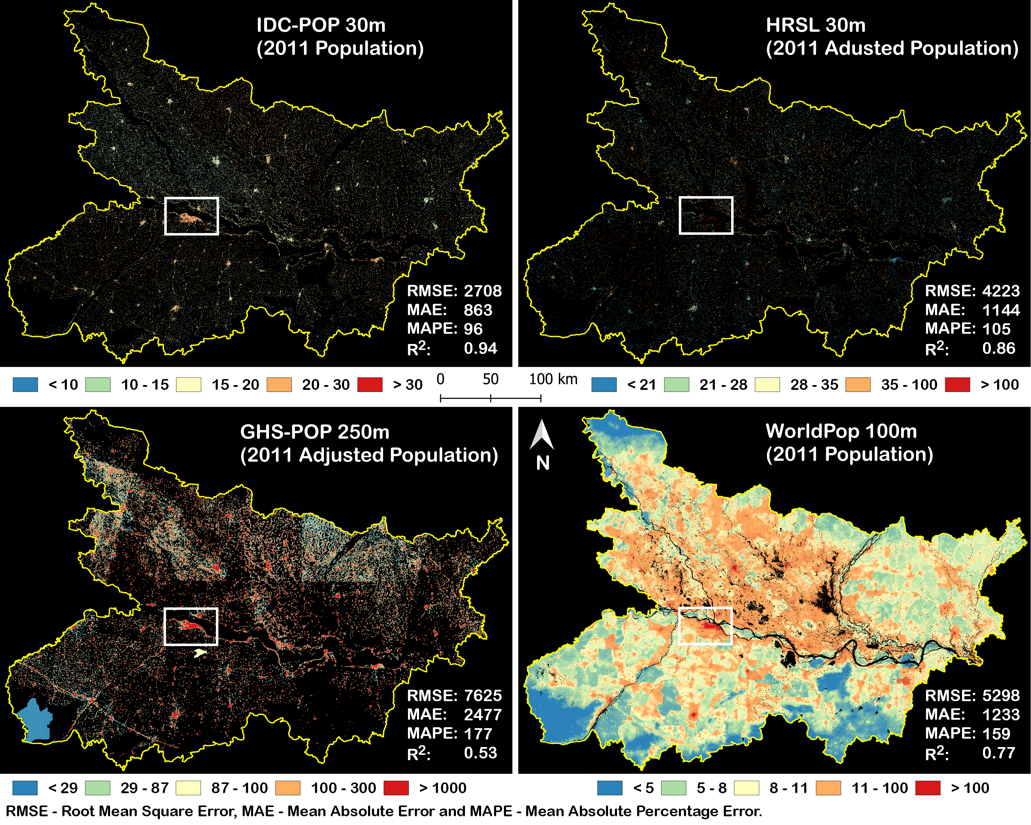}}
\caption{Accuracy comparison of IDC-POP, HRSL, GHS-POP and WorldPop gridded population datasets for Bihar.}
\label{fig:fig10}
\end{figure}

Besides quantitative accuracy assessment at the village/town level, it is also important to ensure that within these towns and villages, people are allocated to potentially habitable areas and not to water bodies, agricultural land or other such areas where people are unlikely to reside. To visually assess this, we show state level maps of all four gridded population datasets for Bihar in Figure \ref{fig:fig10}. 

\end{multicols}

\vspace{-0.25cm}

\newpage

\begin{figure}[H]
\centering
\resizebox{0.95\textwidth}{!}{\includegraphics{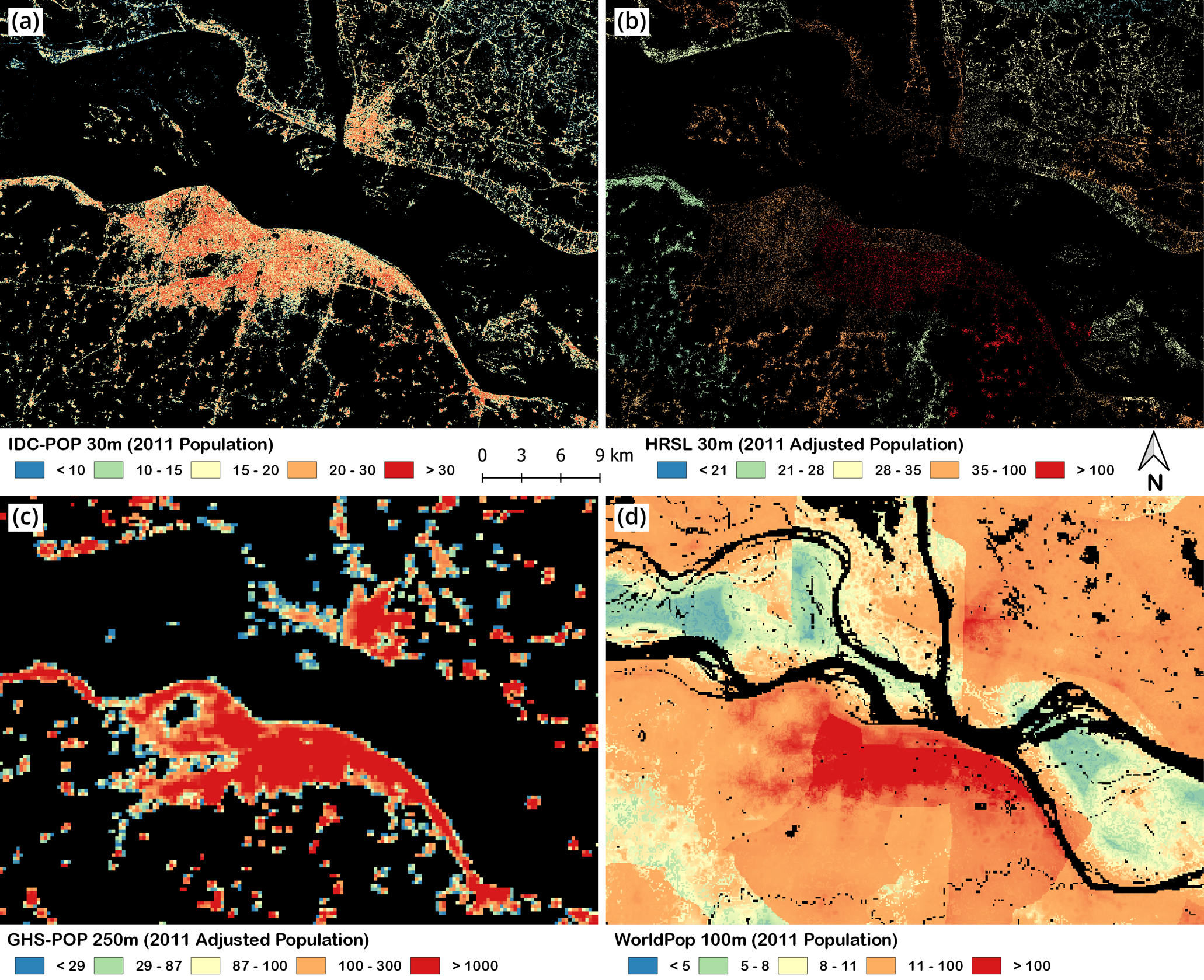}}
\caption{Comparison of population maps for the region around Patna, Bihar (demarcated by the white rectangle in Fig. \ref{fig:fig10}).}
\label{fig:fig11}

\vspace{0.5cm}

\resizebox{0.95\textwidth}{!}{\includegraphics{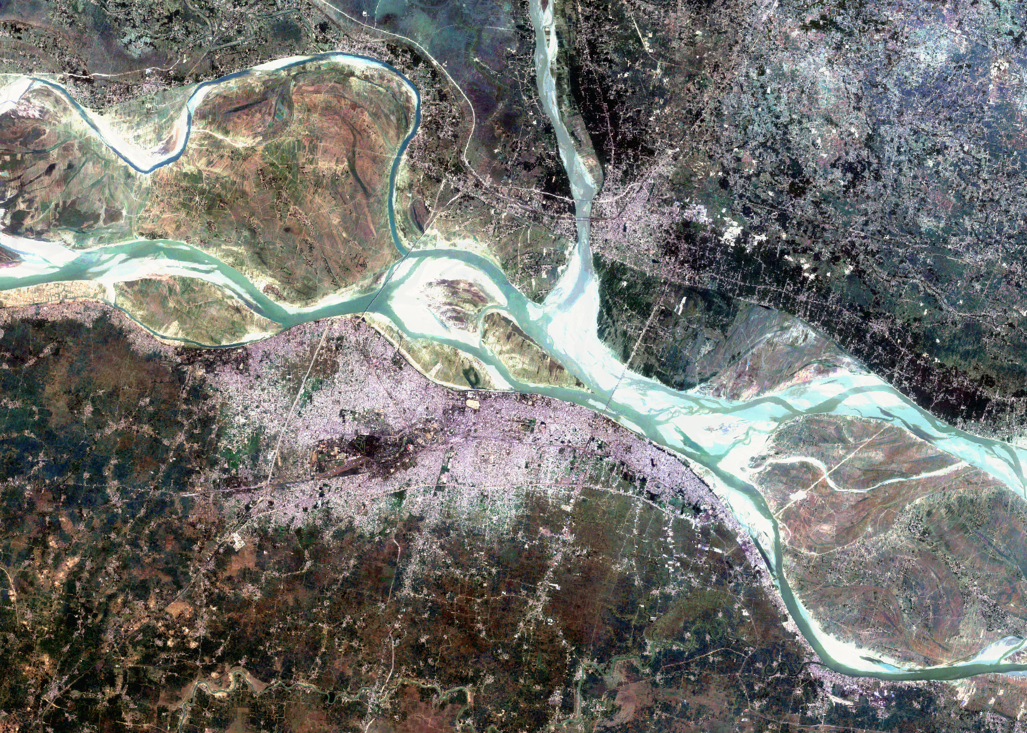}}
\caption{Sentinel-2 MSI (2018) imagery for the region around Patna, Bihar shown in Figure \ref{fig:fig11} (demarcated by the white rectangle in Fig. \ref{fig:fig10}).}
\label{fig:fig12}
\end{figure}

\begin{multicols}{2}

Figure \ref{fig:fig11} shows the area within the white rectangle shown in Figure \ref{fig:fig10}, which is the region around Patna, the capital of Bihar. Figure \ref{fig:fig12} provides Sentinel-2 imagery of this same area for comparison.

As evident from these figures, the IDC-POP and HRSL maps do reasonably well in allocating people to built-up cells. While these may include roads and paved areas, in comparison with GHS-POP and WorldPop, these two datasets perform better. GHS-POP, because of its 250m resolution, allocates people only to the major built-up patches and misses many of the smaller built-up areas visible in Google Earth imagery. WorldPop on the other hand allocates people everywhere except for the major channels of the Ganga River and few other areas. While WorldPop does have a built-up constrained population dataset, it is available only for 2020 projected population. In addition, the built-up layer used by WorldPop to constrain the population mapping for 2020 has far too many errors to be of use in this analysis.

The HRSL dataset, while better than GHS-POP and WorldPop based on the above metrics for Bihar, has a peculiarity that users need to keep in mind: every 30m built-up cell within a sub-district has the same population value. Therefore, while it is technically a 30m resolution dataset, it does not have any variation from cell to cell within a sub-district. This becomes especially problematic in the case of towns since it is unable to capture any intra-urban heterogeneity (eg: Fig \ref{fig:fig11}b). In comparison, the other datasets, including IDC-POP, show heterogeneity within sub-districts.

\section{Conclusion}
From the state and city-level comparisons of built-up datasets and satellite imagery we conclude that at present, GHS-BUILT-S2 is the best 10m built-up layer for India by a considerable margin. Although the WSF, GHS-BUILT-S2 and ESRI datasets are for 2015, 2018 and 2020 respectively, the underprediction of WSF and overprediction of ESRI, with respect to GHS-BUILT-S2, is not explained by this. It is important that users are aware of the accuracy differences between these datasets and use GHS-BUILT-S2 for any analysis which requires built-up data for the 2015 to 2020 time period for  Indian cities or states. 

From our analysis of gridded population datasets we conclude that, among global gridded datasets, HRSL is currently the best option for India. The approach we have used to generate IDC-POP for Bihar, when applied across other states may provide a national level 30m population dataset for India which is better than HRSL--in terms of village/town level accuracy, accuracy in distributing people to potentially habitable areas within towns and villages and in the ability to capture intra-urban and other fine scale heterogeneity. But this is work in progress at present and we hope to complete this by early 2022.

We strongly recommend that the methods behind all open global land cover or gridded population datasets go through a peer review process prior to public data release \cite[eg:][]{corbane2021, stevens2015}. In addition, the accompanying documentation should provide country or state/province level accuracy assessments. In the absence of such accuracy assessments, for land cover, it will be particularly useful to have an online portal which enables comparison of freely available datasets with satellite imagery such as Sentinel-2, or high resolution Google Earth historic imagery for corresponding year. At least for land cover layers which can be identified through visual inspection of high resolution imagery, this would enable users to rapidly evaluate datasets. The POPGRID collaborative has developed a similar tool for gridded population datasets, which could serve as a model for such efforts \cite{popgrid}.

\section*{Funding Information}

\hspace{-0.6cm}This work was completed with support from the PEAK Urban programme, funded by UKRI’s Global Challenges Research Fund, Grant Ref: ES/P011055/1, and a cloud credits grant from the GEO-AWS Earth Observation Cloud Credits Programme.

\end{multicols}

\end{document}